\documentclass[a4paper,11pt]{article}

\usepackage{pos}
\usepackage{lipsum} 
\usepackage{subcaption}
\usepackage[capitalise]{cleveref}
\usepackage{siunitx}
\usepackage{lineno}
\patchcmd{\thebibliography}
  {\labelsep}
  {\itemsep 0pt \labelsep}
  {}{}

\title{Improvements in the Reconstruction of IceCube Realtime Alerts}

\ShortTitle{Improvements in the Reconstruction of IceCube Realtime Alerts}

\author{The IceCube Collaboration \\{\normalsize \normalfont(a complete list of authors can be found at the end of the proceedings)}\\}

\emailAdd{gsommani@icecube.wisc.edu}
\emailAdd{tyuan@icecube.wisc.edu}

\abstract{

In 2016, the IceCube Neutrino Observatory launched its realtime program.
When a neutrino candidate of likely astrophysical origin is detected, a public alert is issued, typically within one minute.
These alerts allow the astrophysical community to follow up on the region of the sky where the neutrino likely originated.
Initially, the system issued around six track-signature alerts per year, with a highlight being IceCube-170922A, which was later associated with the flaring blazar TXS 0506+056.
Since 2019, IceCube has expanded the selection criteria for neutrino candidates, increasing the track alert rate to around 30 per year with additional alerts for cascade signatures of probable astrophysical origin implemented in 2020.
This work describes several improvements in the reconstruction of track alerts, which were introduced into the realtime stream in 2024, and details the two algorithms that are alternately used for reconstruction, depending on the reconstructed muon energy.
The improvements result in more precise directional localizations, with a factor of 5 (4) reduction in the 50\% (90\%) contour area.
Systematic errors affecting the reconstructions, such as the ice model or the geometry of the detector, have also been investigated to ensure statistical coverage.
In addition to its application in the realtime stream, the improvements described here are also being applied to historical alerts with an updated catalog of track alerts forthcoming.

\vspace{4mm}

{\bfseries Corresponding authors:}
Giacomo Sommani$^{1*}$, Tianlu Yuan$^{2}$\\
{$^{1}$ \itshape Ruhr-Universität Bochum}\\
{$^{2}$ \itshape Dept.~of Physics and Wisconsin IceCube Particle Astrophysics Center, University of Wisconsin–Madison, USA}\\[4mm]
$^*$ Presenter
}

\ConferenceLogo{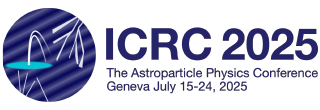}

\FullConference{39th International Cosmic Ray Conference (ICRC2025)\\
 15–24 July 2025\\
Geneva, Switzerland\\}

\begin{document}

\maketitle

\vspace{-3mm}
\section{Introduction}\label{intro}
\vspace{-3mm}

The IceCube Neutrino Observatory is located at the South Pole and consists of a cubic kilometer of glacial ice instrumented with digital optical modules~(DOMs), each containing a downward-facing photomultiplier tube (PMT)~\cite{Aartsen:2016nxy}.
When neutrinos interact with Antarctic ice, they produce charged particles that induce Cherenkov photons, which are detectable by the PMTs.
The light-emission signatures can be classified into two event morphologies: tracks and cascades.
Track-like events are produced by muons, which originate from cosmic-ray showers or charged-current (CC) interactions of muon neutrinos.
Cascade-like events can result from neutral-current interactions of all-flavor neutrinos or from CC interactions of electron and tau neutrinos.
Track signatures typically offer good reconstruction of the direction but imprecise energy estimation.
Cascade signatures, on the other hand, allow for a precise estimate of the neutrino energy, but offer less precise directional reconstruction.

In 2013, the IceCube collaboration announced the detection of a diffuse astrophysical neutrino flux~\cite{neutrinoflux}, but the sources of these neutrinos are still largely unknown.
To investigate the possibility that powerful, transient phenomena produce neutrinos, the IceCube Realtime Alert System was established in 2016~\cite{Aartsen:2016lmt}.
This system allows the astrophysical community to rapidly follow up on the region of the sky where neutrinos that are likely of astrophysical origin were detected.
Initially, this system issued an average of six track-signature alerts per year.
One was IceCube-170922A, later associated with the flaring blazar TXS 0506+056~\cite{txs_coincidence}.
In 2019, the selection criteria were revised and expanded, increasing the track alert rate to an average of 30 per year~\cite{Blaufuss:2019fgv}.
Moreover, in 2020, additional alerts for cascade signatures were implemented\footnote{\url{https://gcn.gsfc.nasa.gov/doc/High_Energy_Neutrino_Cascade_Alerts.pdf}}.

Each track alert consists of a first GCN Notice\footnote{\url{https://gcn.nasa.gov/notices}}~(Revision 0) within a minute of the event, followed by a GCN Circular within a few hours containing updated directional information\footnote{\url{https://gcn.nasa.gov/circulars}}.
As soon as the GCN Circular is released, the GCN Notice is updated accordingly~(Revision 1).
The Revision 0 contains the direction reconstructed with a fast algorithm, referred to as \textit{SplineMPE}~\cite{Abbasi_2021}, while the Revision 1 provides results from a more sophisticated and computationally expensive reconstruction based on and referred to as \textit{Millipede}~\cite{Aartsen:2013vja}.
In September 2024~\cite{gcn_update}, \textit{Millipede} has been replaced by a combination of two updated reconstruction methods, \textit{Millipede Wilks} and \textit{SplineMPE with likelihood scan}.
This contribution aims to detail how the two methods were combined to ensure a minimal size of the contour areas on which follow-up observations would be performed, while also providing consistent and well-characterized statistical coverage.

Sec.~\ref{sec:challenges} briefly describes the likelihood scan method and covers issues arising in the original Millipede reconstruction.
Sections \ref{wilks} and \ref{splinempe} describe the two reconstruction methods involved in the update.
Section \ref{recoquality} details the coverage and robustness of both methods, and Sec.~\ref{hedled} discusses how to combine the two approaches for optimal coverage and precision across all track alerts.
Section \ref{conclusion} summarizes the improvements achieved.

\vspace{-3mm}
\section{Challenges in the original Millipede reconstruction}\label{sec:challenges}
\vspace{-3mm}
The original Millipede algorithm that had been employed in the follow-up reconstruction of IceCube realtime alerts fits to stochastic energy losses along the high-energy muon track~\cite{Aartsen:2013vja}. These energy losses appear as electromagnetic (EM) or hadronic showers, and therefore are each modeled assuming a cascade hypothesis. The track can thus be segmented into a series of multiple cascades, with identical directions corresponding to the direction of the track itself. In order to obtain full-sky, profile log-likelihood maps, we use the Hierarchical Equal Area isoLatitude Pixelization (HEALPix) framework~\cite{Gorski_2005} to pixelate the sky into a fine grid and for each fixed direction, profile over other parameters such as the position, in order to construct a map over the sky giving the likelihood of each pixel's direction. This scan was then used to infer the 50\% and 90\% uncertainty regions for the alerts.

A previous study found that the error contours derived with Millipede did not match the expected coverage~\cite{gualda2021studies}. In particular, the per-event cumulative distribution functions (CDF) of  $2(l^* - \hat{l})$, where $l^*$ and $\hat{l}$ are the log-likelihoods at the best-fit and true directions, did not converge (c.f.~left panel of Fig.~3, Ref.~\cite{millipede2023}). Thus, Wilks' theorem was previously not applicable and when applied often resulted in extremely small contours that did not cover the true direction at the specified levels. Instead, the conversion of the IceCat-1 log-likelihood map into confidence regions was performed based on resimulations of a single event~\cite{panstarrs}. This corresponded to using $2(l^* - \hat{l})= 22.2~ (64.2)$ as the \SI{50}{\%} (\SI{90}{\%}) contour levels, much larger than those obtained from $\chi^2(k=2)$. For the majority of events, these values substantially enlarged the obtained contours but, since the CDFs did not converge, the contours obtained do not accurately represent the respective confidence regions for the general population of realtime alerts.

The issue motivated a search for alternative solutions for the realtime track alerts.
A simulated data set of neutrino alert events heterogeneous in energies and directions, known as \textit{the realtime benchmark simulations}, was introduced~\cite{splinempe2023}.
This dataset consists of 100 simulated events that would have triggered an alert if they had been observed in real data.
Each of these events was re-simulated 100 times by fixing the muon energy losses along the track and performing the propagation of the Cherenkov photons, while varying the parameters of the ice model with the simulation tool \textit{SnowStorm}~\cite{Aartsen_2019}. Due to variations in optical properties of the Antarctic ice from \textit{SnowStorm}, and inherent stochasticity in Cherenkov photon emissions, repeated simulation of a single neutrino event leads to slightly different detector observations. These simulated events were analyzed using an improved version of Millipede referred to as \textit{Millipede Wilks}~(Sec.~\ref{wilks} and~\cite{millipede2023}) and with SplineMPE in a new implementation with likelihood scan~(Sec.~\ref{splinempe} and~\cite{splinempe2023}).
Both approaches resulted in significant improvements compared to the original Millipede method and led to an update of the reconstruction for realtime track alerts in September 2024~\cite{gcn_update}.

\vspace{-3mm}
\section{Millipede Wilks}\label{wilks}
\vspace{-3mm}

Millipede Wilks is an algorithm that can be used to reconstruct high-energy tracks using the same segmented approach as the original Millipede routine, while incorporating recent updated modeling of the in-ice particle shower and various improved minimization settings~\cite{millipede2023,IceCube:2024csv}. Since the inception of the realtime program, a tremendous amount of progress has been made in the understanding of the glacial ice. These include a model of the observed anisotropy of arrival photons from calibration LED devices with ice-crystal birefringence and the polycrystalline structure of the ice~\cite{tc-18-75-2024}, and an improved mapping of ice layer undulations~\cite{IceCube:2023qua}. The expected photoelectron yields from in-ice particle showers is strongly dependent on ice properties, and these effects have been taken into account for reconstruction purposes with an improved cascade model~\cite{IceCube:2024csv}. The model can be naturally adapted for the reconstruction of tracks that trigger a realtime alert, and as detailed in Ref.~\cite{millipede2023} has been shown through to improve the statistical coverage of the reported contours.

\begin{figure*}[hbt]
\centering
\includegraphics[width=0.365\textwidth]{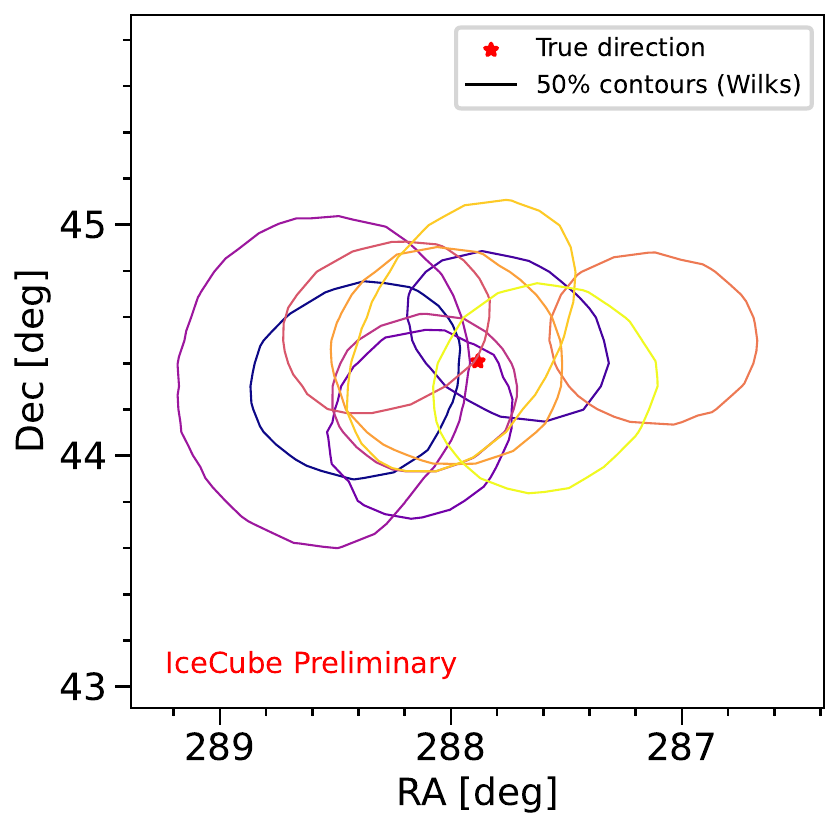} 
\includegraphics[width=0.49\textwidth]{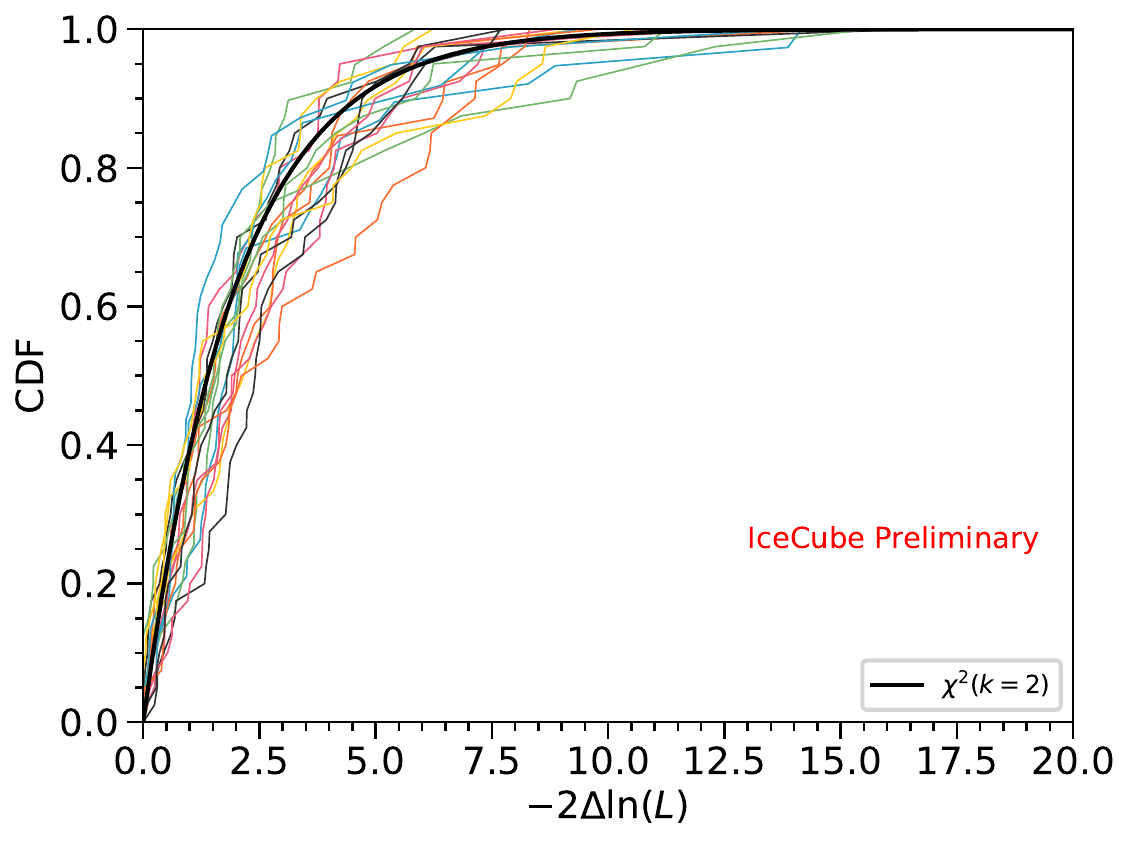}
\caption{Both panels show results from the Millipede Wilks reconstruction. In the left panel, each colored line represents 50\% contours as evaluated by applying Wilks' theorem to a full-sky likelihood scan of ten resimulations of the same underlying event. The red star indicates the true arrival direction, and taken together the figure illustrates how coverage is calculated. In the right panel, each colored line shows the cumulative distribution function (CDF) of $-2 \Delta \ln(L) \equiv 2(l^*-\hat{l})$, where $l^*$ and $\hat{l}$ are the negative log-likelihoods at the true and best-fit directions, respectively, compiled from resimulations of the same underlying event. Ideally, the CDFs of all events would converge towards the black line, corresponding to a chi-2 distribution with $k=2$ degrees of freedom, thus ensuring accurate interpretation of the statistical coverage.}
\label{fig:coverage}
\end{figure*}

Since ICRC 2023, further improvements have been made in the modeling of photon arrival time distributions from in-ice particle showers~\cite{IceCube:2024csv}. The results described in this proceeding include those updates. Using the MC sample of resimulated alert-like events, we evaluated the coverage and systematic robustness of the Millipede Wilks reconstruction. To illustrate how per-event coverage is evaluated, the left panel of \cref{fig:coverage} shows the \SI{50}{\%} likelihood contours obtained by applying Wilks' theorem to the log-likelihood map over the sky. The true direction is indicated by the red star, and for accurate statistical coverage we expect it to lie within the \SI{50}{\%} contours about half the time. To translate this into a general statement across how Wilksian the log-likelihood space is, the CDF of twice the difference between the log-likelihood at the true direction and the best-fit direction, $2(l^* - \hat{l})$, can be compiled on a per-event basis. The result is shown in the right panel of \cref{fig:coverage}, along with the CDF of a chi-2 distribution with $k=2$ degrees of freedom, for the same subset of events as shown in Fig.~3 of Ref.~\cite{millipede2023}. Each colored line is constructed using resimulations of the same event to construct the CDF of $2(l^*-\hat{l})$ across the different resimulations. Ideally, a convergence of the CDFs to the chi-2 CDF would support the validity of Wilks' theorem, and generally convergence to any line even if not $\chi^2(k=2)$ would still allow for the construction of a robust mapping between log-likelihood levels and confidence intervals with correct coverage. A direct comparison of the right panel to Fig.~3 of Ref.~\cite{millipede2023} highlights the convergence improvements since that time, and also illustrates that Wilks' theorem is now broadly more applicable than before.

\vspace{-3mm}
\section{SplineMPE with likelihood scan}\label{splinempe}
\vspace{-3mm}

Most IceCube analyses looking for neutrino sources use the reconstruction method SplineMPE~\cite{Abbasi_2021}.
Unlike Millipede, SplineMPE assumes that light deposition in the detector is continuous.
Under this assumption, the light is not induced by the stochastic energy losses (Sec.~\ref{sec:challenges} and~\cite{Aartsen:2013vja}), but it is continuous and uniform along the whole track.
This scenario is less realistic than the stochastic one, and the higher the muon energy, the less accurate the hypothesis becomes.
Nonetheless, this assumption simplifies the likelihood used to recover the original direction of the neutrino-induced muon, as per each DOM, only two information are required: the arrival time of the first-detected photon and the total number of photons detected by that DOM.
The small amount of data used for the reconstruction makes SplineMPE intrinsically very robust against systematic uncertainties in the ice modeling~\cite{splinempe2023}.

In a previous study~\cite{splinempe2023}, a likelihood scan identical to the one used with Millipede and Millipede Wilks (Sec.~\ref{sec:challenges}) was applied to SplineMPE to reduce minimization issues and provide uncertainty contours better resembling the unique features of the event.
The new method was then tested using alert-like events with promising results.
First, the method demonstrated high precision, with $\sim90\%$ of the realtime benchmark simulations (Sec.~\ref{intro}) having the true direction within~$0.5^\circ$ from the reconstructed one.
Second, the coverage of the uncertainty contours obtained using Wilks' theorem was significantly improved compared to the original Millipede.

\vspace{-3mm}
\section{Coverage and robustness of the two methods}\label{recoquality}
\vspace{-3mm}

\begin{figure*}[t]
\centering
\begin{subfigure}{0.45\textwidth}
\includegraphics[width=\linewidth]{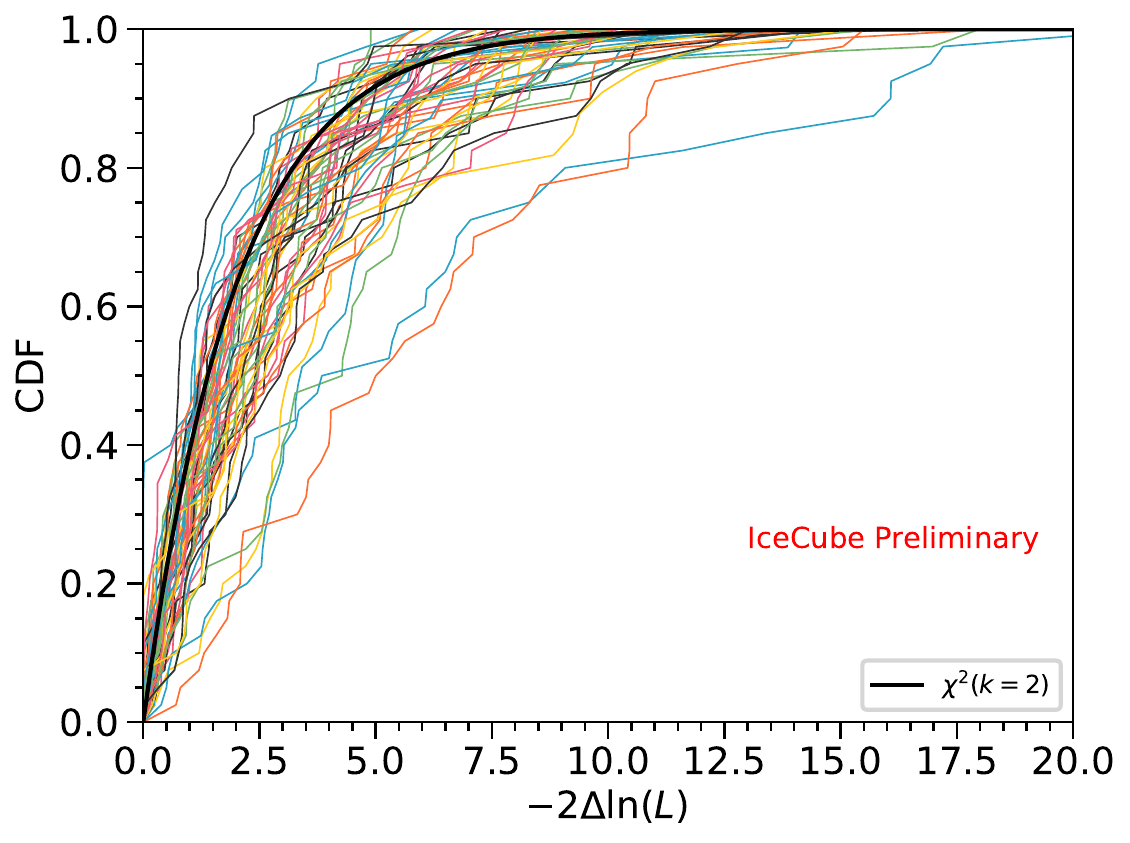}
\caption{Millipede Wilks}
\end{subfigure}
\begin{subfigure}{0.45\textwidth}
\includegraphics[width=\linewidth]{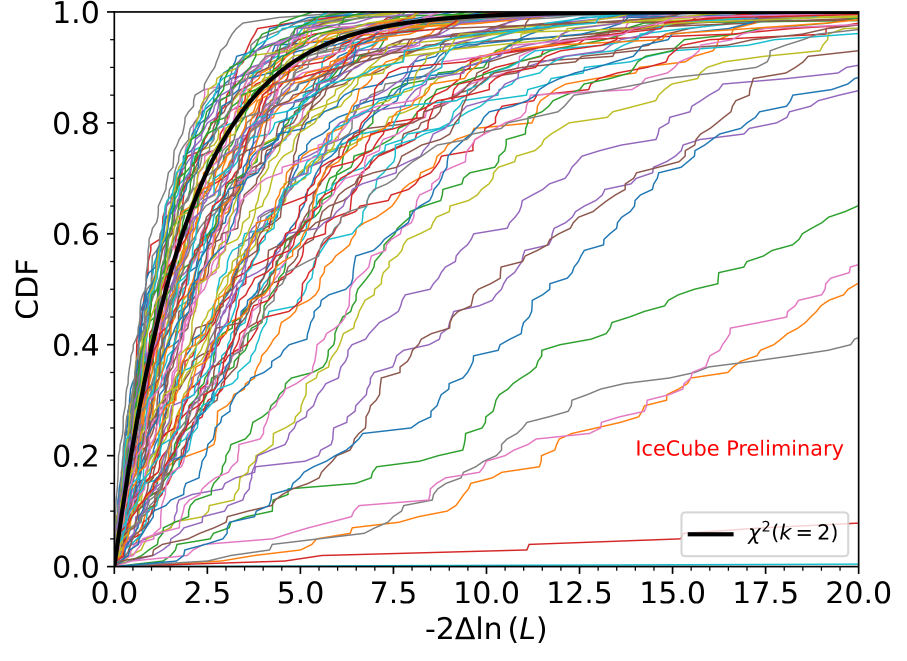}
\caption{SplineMPE with likelihood scan}
\end{subfigure}
\caption{
In both panels, each colored line shows the cumulative distribution function (CDF) of $2(l^*-\hat{l})$, where $l^*$ and $\hat{l}$ are the log-likelihoods at the best-fit and true directions, respectively, compiled from resimulations of the same underlying event.
Each colored line corresponds to one of the realtime benchmark simulations~\cite{splinempe2023}, each resimulated 100 times.
Ideally, the CDFs of all events would converge towards the black line, corresponding to a chi-2 distribution with $k=2$ degrees of freedom, thus ensuring accurate interpretation of the statistical coverage.
On the left panel, the results obtained with Millipede Wilks~(Sec.~\ref{wilks}) are overall in good agreement with Wilks' expectations.
On the right panel, the results obtained with SplineMPE with likelihood scan~(Sec.~\ref{splinempe}) but without the application of an angular error floor (see~Sec.~\ref{recoquality}), with some events not in agreement with the expectations.
}
\label{fig:llh_distr}
\end{figure*}

In this work, we tested the coverage of both Millipede Wilks~(Sec.~\ref{wilks}) and SplineMPE with likelihood scan~(Sec.~\ref{splinempe}) using the realtime benchmark simulations~(Sec.~\ref{sec:challenges} and \cite{splinempe2023}).
Figure~\ref{fig:llh_distr} shows the cumulative density functions (CDFs) of the likelihood ratio, i.e., $2(l^* - \hat{l})$, where $l^*$ and $\hat{l}$ are the log-likelihoods at the true and best-fit directions, for both methods.
Millipede Wilks has overall better statistical coverage, while SplineMPE with the likelihood scan has many events far away from the expected CDF distribution of Wilks' theorem.

\begin{figure}[hbt]
\centering
\begin{subfigure}{0.43\linewidth}
\includegraphics[width=\linewidth]{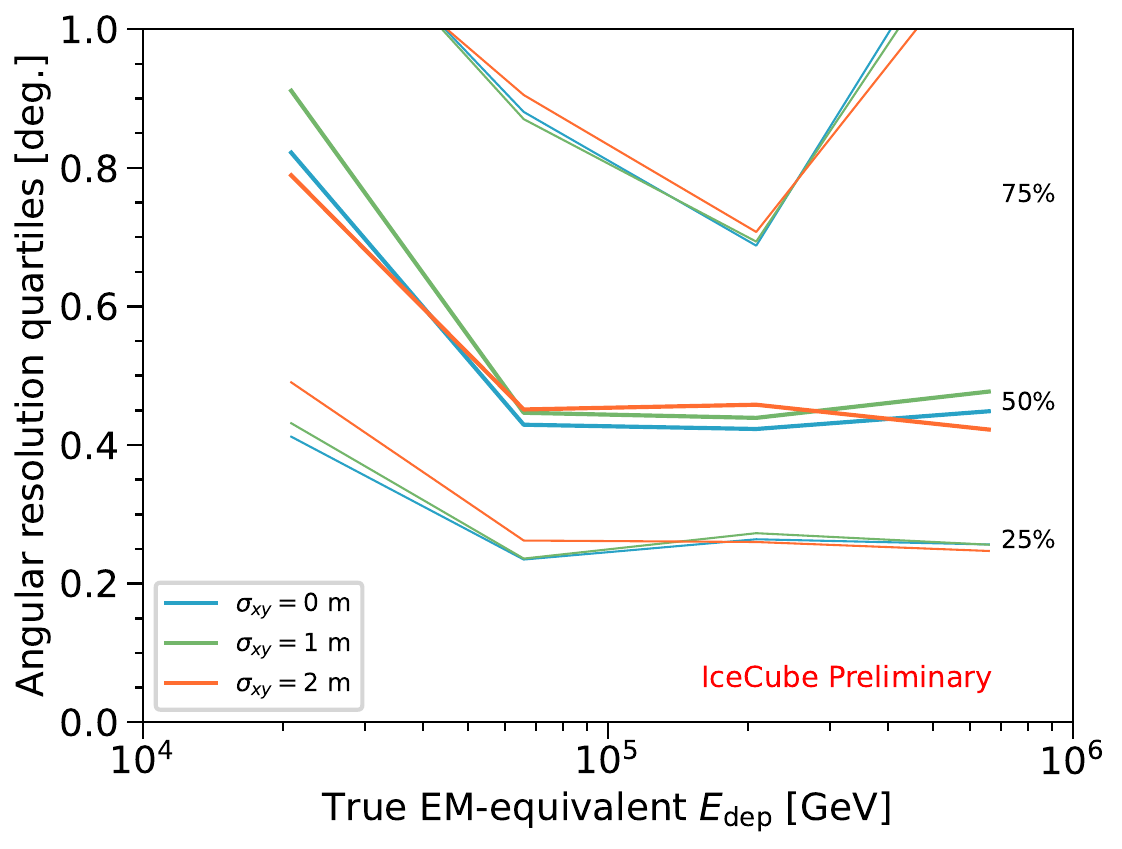}
\caption{Millipede Wilks}
\end{subfigure}
\begin{subfigure}{0.43\linewidth}
\includegraphics[width=\linewidth]{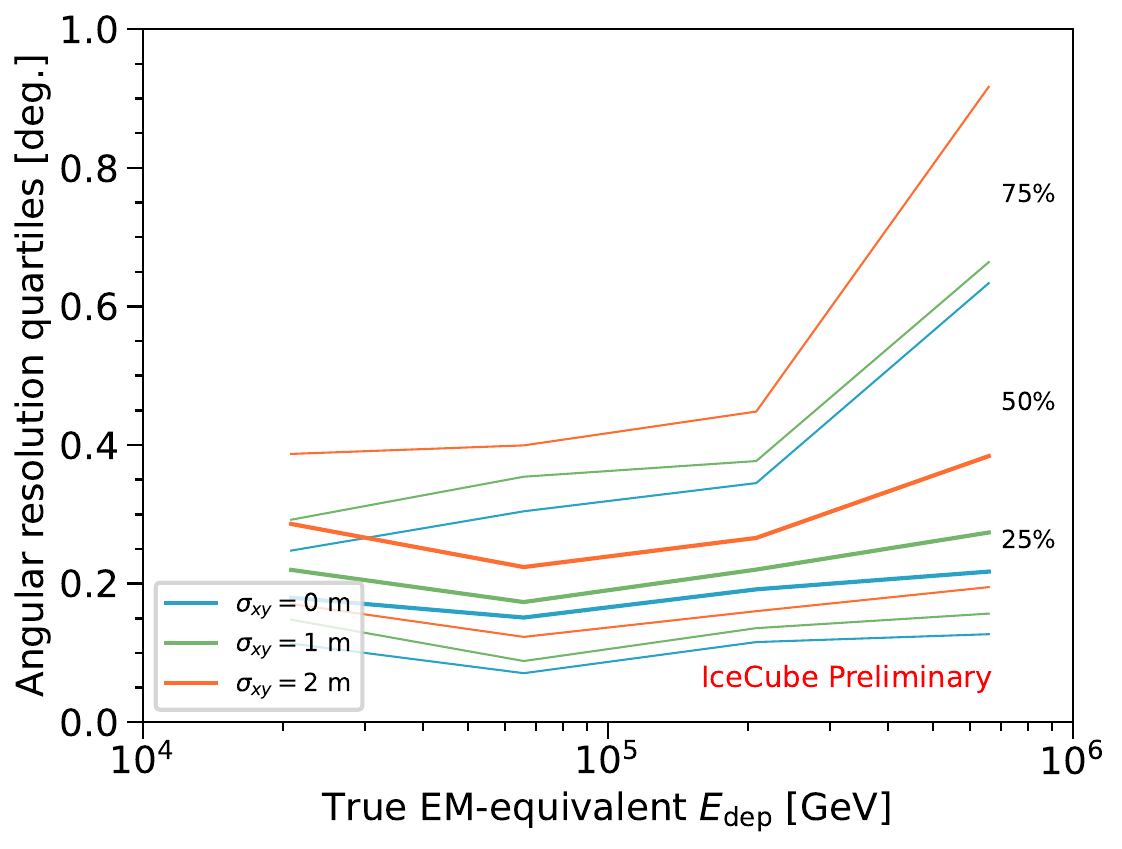}
\caption{SplineMPE with likelihood scan}
\end{subfigure}
\caption{
Impact of geometry variations on the Millipede Wilks (left) and SplineMPE with likelihood scan (right).
The figures show the impact on the angular resolution quartiles (25/50/75) on a subset of the realtime benchmark simulations.
On SplineMPE with likelihood scan, the impact of the geometry variations is much higher than on Millipede Wilks.
}
\label{fig:geometry}
\end{figure}

In this work, we also tested both methods against systematic uncertainties in the detector geometry.
In the resimulations of the realtime benchmark simulations, each DOM position was varied in the horizontal plane following a Gaussian distribution with a standard deviation equal to 1 and 2 m.
Figure~\ref{fig:geometry} shows the results of these tests.
While Millipede Wilks seems almost unaffected by these variations, SplineMPE with likelihood scan shows a worsening of its precision of the order of 0.1 to 0.2 degrees.
To take the effect of the systematic uncertainties in the geometry into account for SplineMPE, its likelihood scans are convoluted with a bi-dimensional symmetric Gaussian distribution with a standard deviation of 0.2 degrees.
This convolution will be further referred to as \textit{angular error floor}.

\vspace{-3mm}
\section{Low and High Energy Deposition Events}\label{hedled}
\vspace{-3mm}

\begin{figure*}[hbt]
\centering
\begin{subfigure}{0.43\linewidth}
\includegraphics[width=\linewidth]{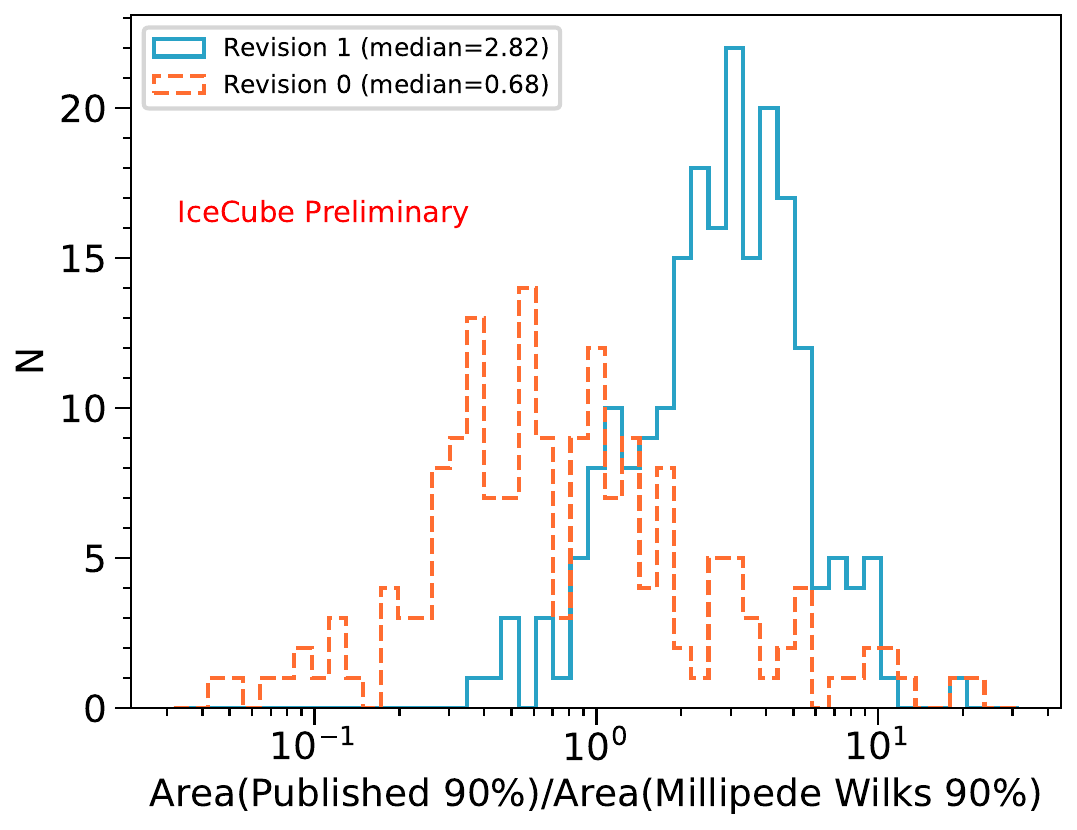}
\caption{Millipede Wilks}
\end{subfigure}
\begin{subfigure}{0.43\linewidth}
\includegraphics[width=\linewidth]{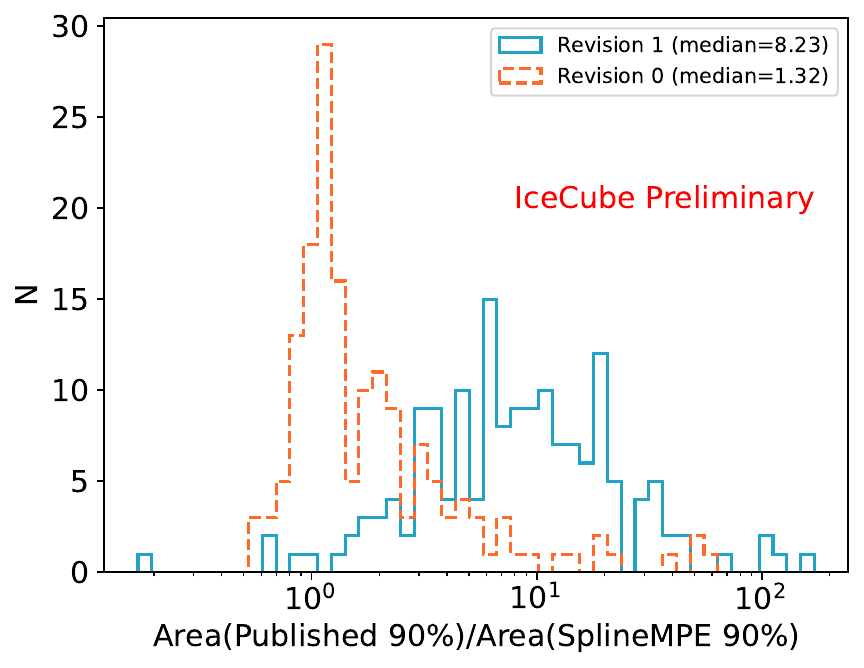}
\caption{SplineMPE with likelihood scan}
\end{subfigure}
\caption{Comparison of 90\% areas between Millipede Wilks (left) and SplineMPE with likelihood scan (right) against areas of the original Millipede, Revision 1 (blue) and initial GCN, Revision 0 (orange) for events in IceCat-1. The 90\% areas are computed for each published revision, and their ratio is taken with respect to the 90\% area obtained with Millipede Wilks or SplineMPE with likelihood scan.}
\label{fig:areas}
\end{figure*}

\begin{figure*}[hbt]
\centering
\includegraphics[width=0.43\textwidth]{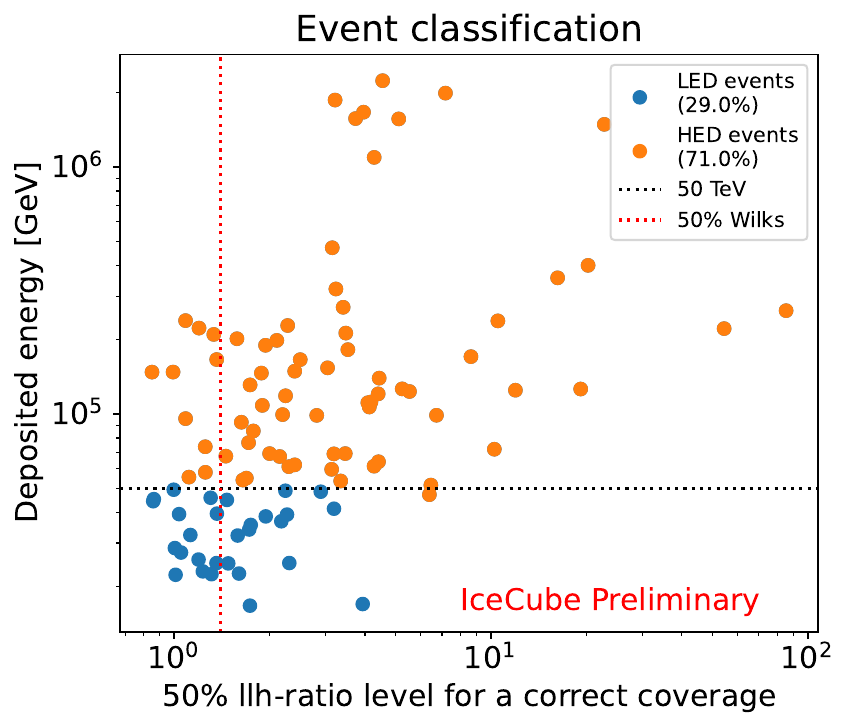}
\includegraphics[width=0.45\textwidth]{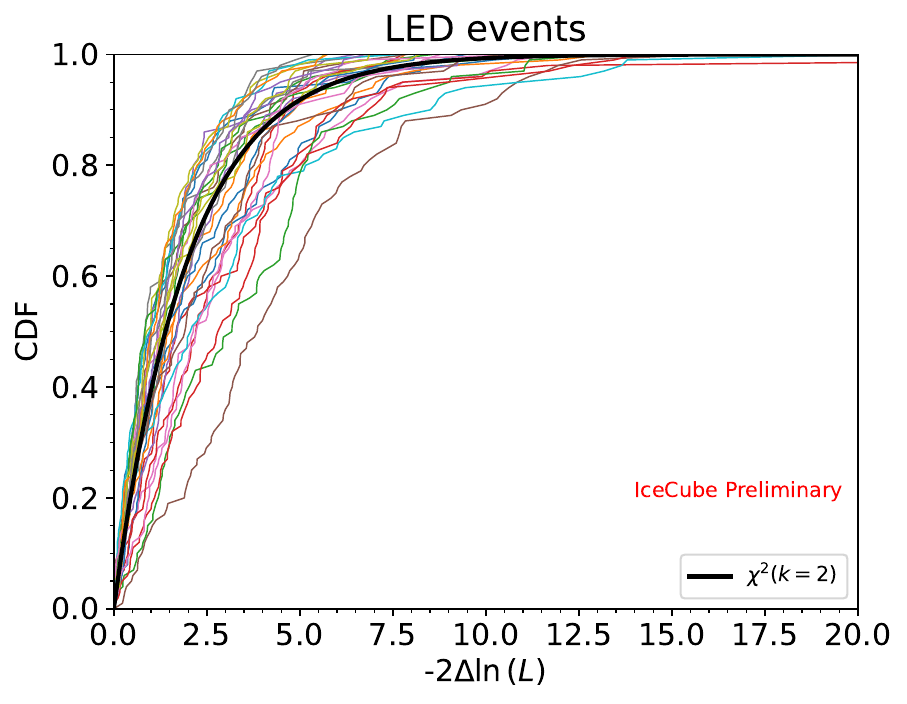}
\caption{
In the left panel, classification of low energy deposition (LED) and high energy deposition (HED) events.
On the y-axis is the energy deposited in the detector by the muon.
On the x-axis is the median level of log-likelihood ratio for each realtime benchmark simulation (resimulated 100 times), from SplineMPE with likelihood scan.
This median level also represents the llh-ratio level at which there would be a correct 50\% coverage for that event.
The red vertical dotted line represents the expected 50\% level from Wilks' Theorem.
The more the dots are further right, the worse the reconstruction.
The black horizontal dotted line indicates the deposited energy that distinguishes between LED and HED events.
HED events are also all alerts that passed the HESE filter~\cite{Blaufuss:2019fgv}, as shown by the example of the orange dot slightly below the 50 TeV line.
In the right panel, the same plot as the right panel of Fig.~\ref{fig:llh_distr}, but with the results from the SplineMPE likelihood scan applied only to the LED events.
On this subset of events, the likelihood ratio CDFs are more in agreement with the expectations from Wilks' theorem.
}
\label{fig:classification}
\end{figure*}

Millipede Wilks showed in general a good statistical coverage (Sec.~\ref{recoquality} and left panel of Fig.~\ref{fig:llh_distr}) over all alert-like events, and better than SplineMPE with likelihood scan.
However, in this work, we also compared the uncertainty areas of the two methods; despite the angular error floor applied, the ones of SplineMPE with likelihood scan are significantly smaller.
Fig~\ref{fig:areas} shows the ratio between the published 90\% areas in the Revisions 0 and 1 for each alert (see Sec.~\ref{intro}) and the new methods Millipede Wilks (left panel) and SplineMPE with likelihood scan (right panel).
From the median values of these ratios, SplineMPE with likelihood scan has contours 2 to 3 times smaller than Millipede Wilks.
Therefore, if SplineMPE with likelihood scan demonstrates good statistical properties on a subset of the realtime alerts, it is to be preferred on this subset over Millipede Wilks, also to facilitate subsequent follow-up observations.
In this work, we studied the statistical coverage of the uncertainty regions of SplineMPE with likelihood scan in relation to the deposited energy in the detector estimated using the methods explained in~\cite{ABBASI2013190} (left panel of Fig.~\ref{fig:classification}).
When the deposited energy is very high, the coverage of the errors of SplineMPE with likelihood scan becomes unpredictable and typically leans towards undercoverage.
On the other hand, for small deposited energies, it seems to fluctuate consistently around the expectations from Wilks' theorem.
The left panel of Fig.~\ref{fig:classification} shows the realtime benchmark simulations (Sec.~\ref{intro}, \cite{splinempe2023}) in a space of deposited energy and log-likelihood ratio level for a 50\% coverage.
Their density in that bi-dimensional space was interpolated using different methods, and 50~TeV was found to be the energy at which, consistently with all methods, more than 10\% of the alerts at that energy would have a 90\% contour (from Wilks' theorem) with less than 50\% of real coverage.
Therefore, 50~TeV was chosen as the energy threshold to distinguish between a subset of events on which SplineMPE with likelihood scan is preferable to Millipede Wilks and the rest of the events on which Millipede Wilks is adopted for its good statistical properties, which is still a major improvement in precision to the old reconstruction of realtime alerts (left panel of Fig.~\ref{fig:areas}).
These two subsets are respectively referred to as \textit{low-energy deposition} (LED) and \textit{high-energy deposition} (HED) events.
The right panel of Figure~\ref{fig:classification} shows the coverage properties of SplineMPE with likelihood scan (and without angular-error floor) on the LED events, demonstrating good performance.
Moreover, the alerts that passed the High-Energy-Starting-Event (HESE) selection~\cite{Blaufuss:2019fgv} are also to be classified as HED events independently of their deposited energy, as the initial hadronic shower induced by the interaction of the neutrino with the glacial ice is contained in the detector.
Therefore, the assumption of SplineMPE of continuous light emission is not realistic for these events.

\vspace{-3mm}
\section{Conclusion}\label{conclusion}
\vspace{-3mm}
The combination of the two reconstruction methods, Millipede Wilks and SplineMPE with likelihood scan, ensures a minimal size of the contour areas on which follow-up observations would be performed, while also providing consistent and well-characterized statistical coverage~(Sec.~\ref{hedled}).
While the old catalog of IceCube realtime alerts, namely IceCat-1~\cite{IceCube:2023agq}, was reconstructed with the original Millipede, an updated catalog, IceCat-2, is now being processed~\cite{icecat2}.
Overall, the areas in IceCat-2 shrink by a factor between 4 and 5~\cite{icecat2}.
Further improvements in the reconstruction of realtime track alerts are still possible, and new possibilities are being explored~\cite{Glusenkamp:2020gtr}.

\vspace{-3mm}
\begingroup
\footnotesize
\bibliographystyle{ICRC}
\setlength{\bibsep}{1pt}
\bibliography{references}

\providecommand{\href}[2]{#2}\begingroup\raggedright\begin{thebibliography}{10}

\bibitem{Aartsen:2016nxy}
{\bfseries IceCube} Collaboration, M.~G. Aartsen {\em et~al.}, \href{http://dx.doi.org/10.1088/1748-0221/12/03/P03012}{{\em JINST} {\bfseries 12} no.~03, (2017) P03012}.

\bibitem{neutrinoflux}
{\bfseries IceCube} Collaboration, M.~G. Aartsen {\em et~al.}, \href{http://dx.doi.org/10.1126/science.1242856}{{\em Sci.} {\bfseries 342} no.~6161, (2013) 1242856}.

\bibitem{Aartsen:2016lmt}
{\bfseries IceCube} Collaboration, M.~G. Aartsen {\em et~al.}, \href{http://dx.doi.org/10.1016/j.astropartphys.2017.05.002}{{\em Astropart. Phys.} {\bfseries 92} (2017) 30--41}.

\bibitem{txs_coincidence}
{\bfseries IceCube} Collaboration, Fermi-LAT {\em et~al.}, \href{http://dx.doi.org/10.1126/science.aat1378}{{\em Sci.} {\bfseries 361} no.~6398, (2018) eeat1378}.

\bibitem{Blaufuss:2019fgv}
{\bfseries IceCube} Collaboration, E.~Blaufuss, T.~Kintscher, L.~Lu, and C.~F. Tung, \href{http://dx.doi.org/10.22323/1.358.1021}{{\em PoS} {\bfseries ICRC2019} (2019) 1021}.

\bibitem{Abbasi_2021}
{\bfseries IceCube} Collaboration, R.~Abbasi {\em et~al.}, \href{http://dx.doi.org/10.1088/1748-0221/16/08/P08034}{{\em JINST} {\bfseries 16} no.~08, (2021) P08034}.

\bibitem{Aartsen:2013vja}
{\bfseries IceCube} Collaboration, M.~G. Aartsen {\em et~al.}, \href{http://dx.doi.org/10.1088/1748-0221/9/03/P03009}{{\em JINST} {\bfseries 9} (2014) P03009}.

\bibitem{gcn_update}
{\bfseries IceCube} Collaboration, {\em GCN} {\bfseries 38267} (2024) .

\bibitem{Gorski_2005}
K.~M. Gorski {\em et~al.}, \href{http://dx.doi.org/10.1086/427976}{{\em ApJ} {\bfseries 622} no.~2, (04, 2005) 759–771}.

\bibitem{gualda2021studies}
{\bfseries IceCube} Collaboration, C.~Lagunas~Gualda, Y.~Ashida, A.~Sharma, and H.~Thomas, \href{http://dx.doi.org/10.22323/1.395.1045}{{\em PoS} {\bfseries ICRC2021} (2021) 1045}.

\bibitem{millipede2023}
{\bfseries IceCube} Collaboration, T.~Yuan, \href{http://dx.doi.org/10.22323/1.444.1005}{{\em PoS} {\bfseries ICRC2023} (2023) 1005}.

\bibitem{panstarrs}
{\bfseries {Pan-STARRS}, {IceCube}} Collaboration, E.~Kankare {\em et~al.}, \href{http://dx.doi.org/10.1051/0004-6361/201935171}{{\em {A\&A}} {\bfseries 626} (2019) A117}.

\bibitem{splinempe2023}
{\bfseries IceCube} Collaboration, G.~Sommani, C.~Lagunas~Gualda, and H.~Niederhausen, \href{http://dx.doi.org/10.22323/1.444.1186}{{\em PoS} {\bfseries ICRC2023} (2023) 1186}.

\bibitem{Aartsen_2019}
{\bfseries IceCube} Collaboration, M.~G. Aartsen {\em et~al.}, \href{http://dx.doi.org/10.1088/1475-7516/2019/10/048}{{\em JCAP} {\bfseries 2019} no.~10, (2019) 048--048}.

\bibitem{IceCube:2024csv}
{\bfseries IceCube} Collaboration, R.~Abbasi {\em et~al.}, \href{http://dx.doi.org/10.1088/1748-0221/19/06/P06026}{{\em JINST} {\bfseries 19} no.~06, (2024) P06026}.

\bibitem{tc-18-75-2024}
{\bfseries IceCube} Collaboration, R.~Abbasi {\em et~al.}, \href{http://dx.doi.org/10.5194/tc-18-75-2024}{{\em The Cryosphere} {\bfseries 18} no.~1, (2024) 75--102}.

\bibitem{IceCube:2023qua}
{\bfseries IceCube} Collaboration, R.~Abbasi {\em et~al.}, \href{http://dx.doi.org/10.22323/1.444.0975}{{\em PoS} {\bfseries ICRC2023} (2023) 975}.

\bibitem{ABBASI2013190}
{\bfseries IceCube} Collaboration, R.~Abbasi {\em et~al.}, \href{http://dx.doi.org/https://doi.org/10.1016/j.nima.2012.11.081}{{\em NIM A} {\bfseries 703} (2013) 190--198}.

\bibitem{IceCube:2023agq}
{\bfseries IceCube} Collaboration, R.~Abbasi {\em et~al.}, \href{http://dx.doi.org/10.3847/1538-4365/acfa95}{{\em Astrophys. J. Suppl.} {\bfseries 269} no.~1, (2023) 25}.

\bibitem{icecat2}
{\bfseries IceCube} Collaboration, A.~Zegarelli, A.~Franckowiak, G.~Sommani, N.~Valtonen-Mattila, and T.~Yuan, {\em PoS} {\bfseries ICRC2025} (2025) 1224.

\bibitem{Glusenkamp:2020gtr}
T.~Gl\"usenkamp, \href{http://dx.doi.org/10.1140/epjc/s10052-024-12473-7}{{\em Eur. Phys. J. C} {\bfseries 84} no.~2, (2024) 163}.

\end{thebibliography}\endgroup

%

\clearpage

\section*{Full Author List: IceCube Collaboration}

\scriptsize
\noindent
R. Abbasi$^{16}$,
M. Ackermann$^{63}$,
J. Adams$^{17}$,
S. K. Agarwalla$^{39,\: {\rm a}}$,
J. A. Aguilar$^{10}$,
M. Ahlers$^{21}$,
J.M. Alameddine$^{22}$,
S. Ali$^{35}$,
N. M. Amin$^{43}$,
K. Andeen$^{41}$
% ...
\\
\\
% ...
$^{10}$ Universit{\'e} Libre de Bruxelles, Science Faculty CP230, B-1050 Brussels, Belgium \\
% ...
$^{16}$ Department of Physics, Loyola University Chicago, Chicago, IL 60660, USA \\
$^{17}$ Dept. of Physics and Astronomy, University of Canterbury, Private Bag 4800, Christchurch, New Zealand \\
% ...
$^{21}$ Niels Bohr Institute, University of Copenhagen, DK-2100 Copenhagen, Denmark \\
$^{22}$ Dept. of Physics, TU Dortmund University, D-44221 Dortmund, Germany \\
% ...
$^{35}$ Dept. of Physics and Astronomy, University of Kansas, Lawrence, KS 66045, USA \\
% ...
$^{39}$ Dept. of Physics and Wisconsin IceCube Particle Astrophysics Center, University of Wisconsin{\textemdash}Madison, Madison, WI 53706, USA \\
% ...
$^{41}$ Department of Physics, Marquette University, Milwaukee, WI 53201, USA \\
% ...
$^{43}$ Bartol Research Institute and Dept. of Physics and Astronomy, University of Delaware, Newark, DE 19716, USA \\
% ...
$^{63}$ Deutsches Elektronen-Synchrotron DESY, Platanenallee 6, D-15738 Zeuthen, Germany \\
$^{\rm a}$ also at Institute of Physics, Sachivalaya Marg, Sainik School Post, Bhubaneswar 751005, India \\
% ...

\subsection*{Acknowledgments}

\noindent
The authors gratefully acknowledge the support from the following agencies and institutions:
USA {\textendash} U.S. National Science Foundation-Office of Polar Programs,
U.S. National Science Foundation-Physics Division,
U.S. National Science Foundation-EPSCoR,
U.S. National Science Foundation-Office of Advanced Cyberinfrastructure,
Wisconsin Alumni Research Foundation,
Center for High Throughput Computing (CHTC) at the University of Wisconsin{\textendash}Madison,
Open Science Grid (OSG),
Partnership to Advance Throughput Computing (PATh),
Advanced Cyberinfrastructure Coordination Ecosystem: Services {\&} Support (ACCESS),
Frontera and Ranch computing project at the Texas Advanced Computing Center,
U.S. Department of Energy-National Energy Research Scientific Computing Center,
Particle astrophysics research computing center at the University of Maryland,
Institute for Cyber-Enabled Research at Michigan State University,
Astroparticle physics computational facility at Marquette University,
NVIDIA Corporation,
and Google Cloud Platform;
Belgium {\textendash} Funds for Scientific Research (FRS-FNRS and FWO),
FWO Odysseus and Big Science programmes,
and Belgian Federal Science Policy Office (Belspo);
Germany {\textendash} Bundesministerium f{\"u}r Forschung, Technologie und Raumfahrt (BMFTR),
Deutsche Forschungsgemeinschaft (DFG),
Helmholtz Alliance for Astroparticle Physics (HAP),
Initiative and Networking Fund of the Helmholtz Association,
Deutsches Elektronen Synchrotron (DESY),
and High Performance Computing cluster of the RWTH Aachen;
Sweden {\textendash} Swedish Research Council,
Swedish Polar Research Secretariat,
Swedish National Infrastructure for Computing (SNIC),
and Knut and Alice Wallenberg Foundation;
European Union {\textendash} EGI Advanced Computing for research;
Australia {\textendash} Australian Research Council;
Canada {\textendash} Natural Sciences and Engineering Research Council of Canada,
Calcul Qu{\'e}bec, Compute Ontario, Canada Foundation for Innovation, WestGrid, and Digital Research Alliance of Canada;
Denmark {\textendash} Villum Fonden, Carlsberg Foundation, and European Commission;
New Zealand {\textendash} Marsden Fund;
Japan {\textendash} Japan Society for Promotion of Science (JSPS)
and Institute for Global Prominent Research (IGPR) of Chiba University;
Korea {\textendash} National Research Foundation of Korea (NRF);
Switzerland {\textendash} Swiss National Science Foundation (SNSF).

\end{document}